\documentstyle[prl,twocolumn,aps,epsf]{revtex}

\def\epsfig#1#2#3#4
         {
         \epsfysize=#2 \vbox{ \hglue#3 \epsfbox[#4]{#1} }
         }
\def\epsfigrot#1#2#3#4
         {
         \epsfxsize=#2
         \setbox\rotbox=\hbox to #2{\epsfbox[#4]{#1}}
         \vbox{\hglue#3 \rotl\rotbox}
         }
\newbox\rotbox
\input rotate
\begin{document}
\draft
\title{Precursors, aftershocks, criticality and self-organized
criticality}
\author{Y. Huang$^1$,  H. Saleur$^1$, C. G.  Sammis$^2$, D.
Sornette$^{3,4}$}
\address{$^1$ Department of Physics, University of
Southern California, Los-Angeles, CA 90089-0484.}
\address{$^2$ Department of Earth Sciences, University of
Southern California, Los-Angeles, CA 90089-0740.}
\address{$^3$ Department of Earth and Space Sciences
and Institute of Geophysics and Planetary Physics\\ University of
California, Los Angeles, California 90095-1567}
\address{$^4$  Laboratoire de Physique de la Mati\`ere
Condens\'ee, CNRS URA 190\\
Universit\'e des Sciences, B. P. 70, Parc Valrose, 06108
Nice Cedex 2, France}

\date{\today}
\maketitle

\begin{abstract}
We present a simple model of earthquakes on
a pre-existing hierarchical fault network. The system
self-organizes
on long time scales in a stationary state with a power law
Gutenberg-Richter
distribution of earthquake sizes. The largest fault carries irregular
great earthquakes preceded by precursors developing over long time
scales and followed by aftershocks obeying an  Omori's
law. The cumulative energy released by precursors follows a
time-to-failure power law with log-periodic structures, qualifying a
large event as an effective dynamical (depinning) critical point.
Down the hierarchy, smaller earthquakes exhibit the same
phenomenology,
albeit with increasing irregularities.

\end{abstract}
\vskip 0.2cm

\narrowtext

Seismologists model
dynamical rupture propagation with complex friction laws and barriers
\cite{Madariaga} and attempt to ascribe the
earthquake complexity to nonlinear processes and/or
heterogeneities. From a more global point of view, it has  been
suggested
that earthquakes are  somewhat similar to critical points
\cite{Allegre}, and
can be addressed using tools of the renormalization group.
A very broad, but quite ill-defined,  perspective is also available
with the concept of self-organized criticality \cite{Bak,SS}.
These various  points of view model different properties at different
time
scales; it is hard  to see how they relate to each other, and whether
they are part of a unique meaningful ``theory of earthquakes''. A
closely related puzzle is whether criticality and self organized
criticality are compatible.

The present  work  attempts to unify a significant fraction of this
earthquake phenomenology, and to answer this puzzle: we define a
simple model that exhibits
the  self-critical organization of the crust at large time scales,
the critical
nature of large earthquakes and the short-time rupture dynamic
properties.

We start with  the following  ingredients. $(i)$ The faults are
organized in a hierarchical geometrical structure
\cite{hierarchy,Andrews}. We do not address the problem of the
construction of the fault patterns themselves which involves much
larger
times scale ($10^{5-6}$ years) compared to the time scales we
describe
($10^{0-5}$ years). $(ii)$
The tectonic plate is driven at a slow average uniform rate and we
take
into account its heterogeneities and the existence of relaxation
processes
by allowing for fluctuations in the local rate of loading. $(iii)$
When a
threshold is reached, a redistribution occurs on adjacent faults,
with
amplitude controlled by the size of the faults.

\bigskip
\begin{center}
	\begin{minipage}[htbp]{2.0in}
\vbox{
\epsfysize=4cm
\epsfxsize=5cm
\epsffile{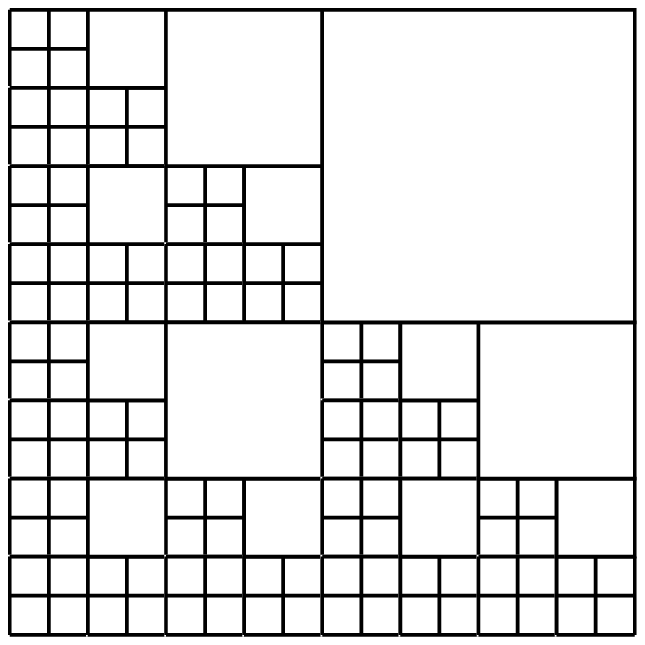}
\begin{figure}
\caption[]{\label{fig1} The fractal cell structure.}
\end{figure}}
\end{minipage}
\end{center}
Our model is an hybridization of the sandpile model \cite{Bak}
and of the fractal automaton \cite{Barriere}.
 As in \cite{Barriere},
the
cell sizes are arranged as a discrete fractal lattice. Each
cell can be viewed as representing the region which is elastically
unloaded when a fault fails. The fractal distribution of cell sizes
then
represents a fractal distribution of fault lengths \cite{hierarchy}.
The
number of fractal generations does not appear to be a crucial
parameter. Most simulations were carried out with $8$ generations,
but
some runs with $12$ generations did not exhibit major
differences. We load the system by dropping particles
at regular intervals (which we use as a clock)
onto the grid at random sites. The addition of a particle is
analogous to
energy loading. The probability that a particle is added to a
particular
cell
is proportional to the area  $A$ of this cell.  Although the exact
cell
(fault domain) to receive the next increment in stress is random, the
entire grid is loaded uniformly at a uniform rate over the long-term.
This represents the long-term uniform
strain at the boundaries between moving tectonic plates. The short
term
random heterogeneity in loading represents heterogeneity in
crustal structure or in upper mantle flow and the associated
relaxation
processes. Each cell becomes unstable when it contains $n \times 4A$
particles,  $n$ being a parameter. It then breaks and redistributes
$4A$
of
its particles to its immediate neighbors. The number of particles
redistributed
to an adjacent cell is proportional to the linear dimension of the
cell.
The $(n-1)\times 4A$ particles that are not redistributed are
considered
as
lost, like the particles which are redistributed outside the grid on
the
plate border. Such energy losses we call ``cooling''. Since the
particles represent energy, the model assumes that a fault fails when
the
stored energy reaches a critical threshold. The key difference
with \cite{Bak} is that the energy must reach the critical level over
the
entire area of a cell before it is allowed to break.  Due to the
fractal
structure, cells of widely different sizes are thus coupled
together, mimicking the multi-scale interactions between faults.

The clock is defined by the particle drops. Cascades are triggered
by the addition of a single  additional particle, {\it i.e.} they
occur
instantaneously (a delay can also be introduced in the aftershock
sequences, see below). At
variance with the rules of \cite{Barriere} and in accord with the
standard sandpile model \cite{Bak}, the size of an earthquake is
determined by the size of the cascade, and is proportional to the
total number of particles cascading. We thus identify the cascade as
complexity in the mainshock, e.g. the linking of fault strands and
segments or even the linking of adjacent faults such as occured in
the
1994 Landers earthquake \cite{Landers}. We define precursors as those
regional  cascades which precede a mainshock and the aftershocks as
the
cascades which follow it. We then use the time scale defined by the
particle
drops to explore the temporal structure of both the foreshock and
aftershock sequences.

The cooling, {\it i.e.} the disappearence  of particles during an
event,
represents a  loss of stored elastic energy due to the earthquake.
Consider for instance an elastic medium under
constant applied shear strain, which suddenly undergoes a rupture in
the form of a dislocation or a crack. The stress field is
redistributed with enhancement at the crack tip as well as screening
at some other places, while both the total shear stress and total
elastic energy decrease. The amount of loss is a
function of the nature of the rupture and of geometry. For instance,
in the Griffith
problem, a crack of length $2c$ is introduced into a rod of
length $L$ and section $\sim L^2$. Under constant strain, the
relative
energy loss ${\delta E \over E}$ incurred by the introduction of the
crack is ${5\alpha c^2 \over L^2}/(1+ {5\alpha c^2 \over L^2})$
\cite{Scholz} with $\alpha$ depending on the geometry and of the
order
of unity. If we take $c$ between $L/4$ and $L/2$, we get ${\delta E
\over
E}$ anywhere between about $0.2-0.6$. This loss corresponds to
friction
on the fault plane, the creation of surface energy in the extensive
crushing which occurs in the fault zone \cite{Andrews}, and in the
radiation of elastic waves \cite{Radiation} whose energy is
ultimately
lost as heat.  Comparing with ${\delta E \over E} =
{n-1 \over n}$ in our model, we see that $n\approx 2$ is not
unreasonable. We take this value for most of our simulations below.
These large uncertainties in cooling are actually not critical - in
fact, provided $n$ is not too close to 1, the results are largely
independent of $n$. The choice $n=1$ of
\cite{Barriere} leading to internal conservation (except at the
boundaries)
does not seem
relevant for  earthquakes.

As in \cite{Barriere}, we identify an event cascading through the
largest cell on the grid as a main shock for our system, {\it i.e.},
the largest regional event. After a transient
depending on the initial conditions, the system self-organizes in a
stationary state with a power law Gutenberg-Richter
distribution of earthquake sizes $P(E) dE \sim E^{-(1.8 \pm 0.1)}
dE$.
This can be called a self-organized critical state, measured by a
statistics
encompassing  many times the largest time intervals between the
largest
earthquakes. We generated about $100$ main shocks on an
$8^{th}$ order fractal structure by adding a total of about $4.6
\cdot 10^7$ particles to the system. The average
number of time units (particle drops) between main shocks is $T=4.6
\cdot 10^5$, with fluctuations of this interval of the order of
$10\%$.
This quasi-periodicity occurs because our  main shocks are
characteristic
earthquakes \cite{Schwartz} which
completely  control the energy redistribution at
large scales.  Their existence  is  not in contradiction with the
self-organized critical state: they  simply arise because of finite
size effects.

For the purpose of comparison with real earthquakes, we choose our
units of time so that $T=100$ years, {\it i.e.} roughly $10^4$ time
units correspond to two years. For most main shocks,
precursory built-up of activity  is clearly
visible, and  lasts $1-2 \cdot 10^5$ time units, or about $20-40$
years. A decaying activity posterior to the main shocks  is also
visible with a lifetime of about $0.1-.2
\cdot 10^5$ time units, {\it i.e.} about $2-4$ years. Quite often,
the
interval between two successive main shocks will not be as quiet, but
is
interrupted by the breaking of one of the second
largest cells.
The time interval between such smaller  shocks is much less
regular than the time interval between main shocks.

We start by discussing aftershocks.
The common wisdom holds that aftershocks involve a time delay
between
the application of the stress and the subsequent rupture.
This delay presumably involves  an intermediate relaxation time,
whose effect could
resemble that of diffusion or visco-elastic processes. In the present
model, the spatial heterogeneity of the loading rate
already  reflects the existence of delay mechanisms.
Inspired by the analysis in terms of critical phenomena (see below),
we then plot the cumulative number of events as a function of time
\cite{foot} starting from an initial date
of about $10^4$ time units posterior to $t_c$ and going backwards in
time.  If
Omori's Law ${dn(t)\over dt}\propto {1\over (t-t_c)^p}$ holds,
this gives $n(t) \sim (t-t_c)^{1-p}$ (the theoretical divergence at
long times for $p<1$ is truncated
by
the existence of  background  seismicity). Fits performed for $15$
of our model aftershock sequences  gave a
distribution
of the exponent $p$ centered around $p=.9$, with small fluctuations,
$p
\in [0.85,1.05]$, in good agreement with the exponent measured
for
earthquakes. These results are independent of the value of
the dissipation parameter $n$, provided $n$ is not too close to $1$.
As $n
\to 1$, in particular when $n=1$, as in \cite{Barriere}, the exponent
$p$
is often near $1$, although it has large fluctuations, and can be
found as
low as $p=0.7$. Simulations on other fractals indicate that
the exponent $p$ is also near unity for different geometry.

The model can be improved by the addition of a  delay mechanism
with a characteristic time $\tau$ for the
avalanche associated with the main shock (this has no qualitative
effect on the statistics of events, nor on the analysis of
precursors). This had the drawback of introducing one additional
parameter, but allows a more natural analysis of the data.
Fortunately, results were found totally insensitive to the value of
$\tau$ over a wide range ($\tau\leq 10^4$) and in perfect agreement
with
Omori's law again.

\bigskip
\begin{center}
\begin{minipage}[htbp]{3.0in}
\vbox{
\epsfysize=5cm
\epsfxsize=6cm
\epsffile{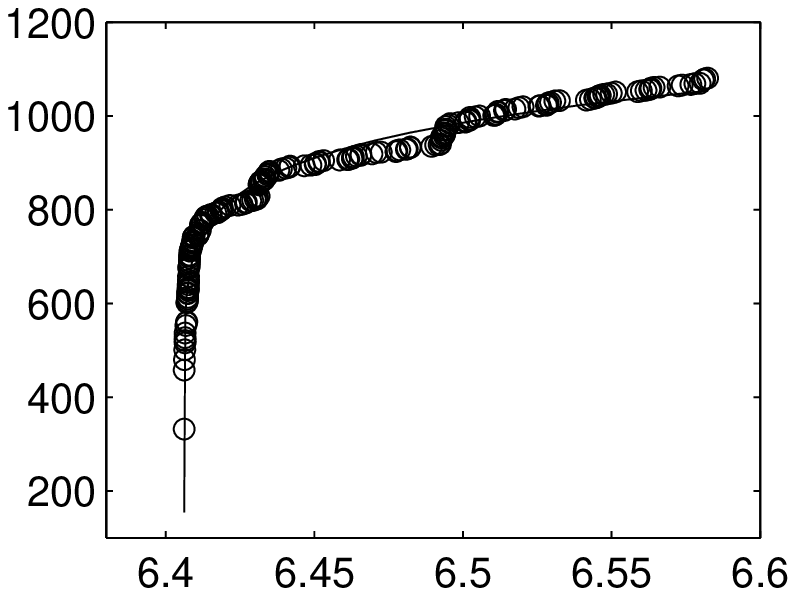}
\begin{figure}
\caption[]{\label{fig2} Cumulative number of afttershocks as  a
function of time (in units of $10^5$ particle drops). The full line
is a
power law with $1-p=.12$.}
\end{figure}}
\end{minipage}
\end{center}

We now turn to precursors. While virtually all large shallow
earthquakes have easily recognizable aftershock sequences, the same
can not be said for foreshocks. In fact, if foreshocks are defined to
occur with the same time and space clustering as aftershocks, then
most
large events do not have a recognizable foreshock sequence
\cite{Jones}
It is only when the time scale is extended to tens of years
and the space to hundreds of kilometers that precursor sequences can
be
recognized \cite{Sykes,Knopoff,logperiodic}. We thus use the term
``precursor'' to distinguish the two definitions. We analyze data in
the same
spirit as in \cite{logperiodic,logperiodic2}: we plot the cumulative
Benioff strain $\epsilon(t)$ (square root of the energy release in an
event) as a function of time, starting  back in time within a range
$1-3
\cdot 10^5$ ($20-60$ years) prior to the main shock. We use a lower
cut-off such as to exclude the crowd of small background events and
test
for various cut-off values. The conclusion is, in most cases, a
reasonably clear evidence for a power law behavior, decorated by log
periodic oscillations \cite{logperiodic,logperiodic2}
\begin{equation}
\epsilon(t)=A - B
(t_c-t)^m\left[1+C\cos\omega\ln(t_c-t)\right].
\label{benioff}
\end{equation}
The fit by this equation of the cumulative foreshock time series for
the
biggest main shock is shown in Figure 3. This example is quite
typical.
Much better fits are sometimes obtained.
Bad fits were rare but sometimes occured.  The improvment of the
$\chi^2$ when using (\ref{benioff}) compared with a pure power law
fit
($C=0$) was always greater than two. The power law in
(\ref{benioff})  is the signature of a critical dynamical behavior,
reminiscent of depinning transition.

\smallskip
\begin{center}
\begin{minipage}[htbp]{3.0in}
\vbox{
\epsfysize=5cm
\epsfxsize=6cm
\epsffile{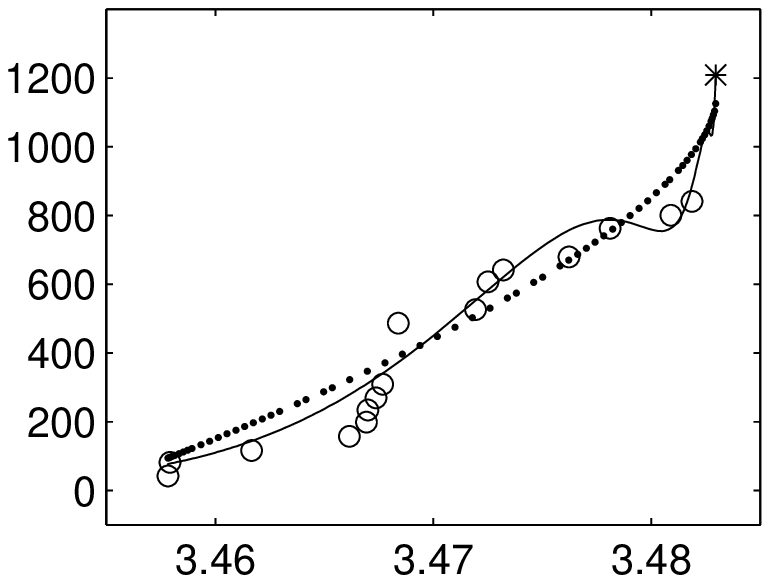}
\begin{figure}
\caption[]{\label{fig3} Plot of Benioff strain before the biggest
shock.  The dotted line  is power law behaviour, full line is power
law decorated with log-periodic oscillations. The time is in units of
$10^7$ particle drops.}
\end{figure}}
\end{minipage}
\end{center}

The log-periodic correction in (\ref{benioff})
reflects the discrete scale invariance (DSI) of the hierarchical
fault network
on which the events occur \cite{logperiodic2} (we observed similar
oscillations decorating the Gutenberg-Richter law). Intriguingly, the
amplitude  $C$ of these oscillations is much bigger than
for equilibrium  statistical mechanics models
 \cite{Ising}: the threshold dynamics in the earthquake model
seems   much more sensitive to DSI. We discuss
below how the value of $\omega$ is related to our fractal
structure.

Values of $m$ and $\omega$
fluctuate more than the exponent $p$ for aftershocks. In $80\%$ of
the
cases however, $m$ was found in the interval $m\in [0.2,0.6]$, in
good
agreement with experimental data \cite{logperiodic,logperiodic2}.
$\omega$ was found in the interval $\omega\in [6,12]$ corresponding
to
$1.7 \leq \lambda \leq 2.8$  where $\omega\equiv 2\pi/\log\lambda$.

These results are not sensitive to the dissipation parameter $n$.
Only
the time scale is modified. In the extreme case of $n=1$ (without
cooling), the seismic activity is much more random, and the rate of
events looks much more constant between main shocks than it does with
cooling: only about $10^4$ time units ($2$ years) before the main
shock
does the cumulative Benioff strain develop  a power law  behavior on
the
approach to the main shock, with properties  that are then similar to
the case $n=2$.

Beside the Benioff strain (\ref{benioff}), we studied the
correlation length $\xi$ defined as the maximum spreading distance of
the
cascades. To get more stable numerical results, we calculated its
integrated value, which  does exhibit a power
law singularity described by
\begin{eqnarray}
\xi &\propto &(t_c-t)^{-\nu_1},\  t<t_c,\
\nu_1\in [0.6,0.8]\nonumber \\
\xi &\propto & (t-t_c)^{-\nu_2},\  t>t_c,\  \nu_2\approx 1.
\label{rtf}
\end{eqnarray}
$\nu_1$ and $\nu_2$ have smaller fluctuations than the exponent $m$
of
the Benioff strain.
Observe that they are different on both sides of
the
critical point, a situation which is known to be possible in
disordered systems in particular. The divergence of $\xi$ in
(\ref{rtf})
confirms the critical point picture. Moreover,  the exponent provides
a relation between DSI in time and DSI in
space.
{}From our fractal geometry, the latter is characterized by $x\to
2x$. Substituting $\xi \to 2\xi$ in
(\ref{rtf}) implies $|t-t_c|\to (\lambda)^{-1} |t-t_c|$
with $\lambda\in [2.4-3.2]$, {\it i.e} $\omega\in [5.4,7.2]$, in
reasonable agreement with what we directly observed.

We now use the time-to-failure law
(\ref{benioff}) to try to forecast the main shock by
fitting it to the ``experimental'' data up to a cutoff time prior
to the main shock, as proposed in \cite{logperiodic}.
Overall, we find that $95\%$ of the main shocks can be predicted
with an uncertainty less than a year, four years in advance. This
must be
compared with the typical fluctuations of about $10$ years of the
time intervals between two main consecutive shocks. There are however
cases where the predictability is much higher, and also extreme cases
where the precursory activity is essentially nonexistent.

We also studied  events occuring on  the second largest cells. The
analysis is slightly more complicated because
two such events can be close in time but well-separated in space. We
therefore restricted our attention to well isolated cases. The
analysis of precursors and aftershocks gives similar results as for
the
foregoing events on largest cells. The only difference is that the
time scales involved are shorter  (roughly by a factor of $10$), and
the
fluctuations in exponents somewhat bigger. This is presumably due to
the fact that
the relative size of the fluctuations of the ``energy'' field with
respect to that of these earthquakes is larger for smaller events due
to
the influence of the earthquakes at the upper levels.

Our results  thus indicate that reasonable intermediate-time
earthquake
prediction may be achievable, as proposed in
\cite{logperiodic,logperiodic2}. From the simulations, there is clear
evidence that the predictabilty depends on the ``temperature'' of the
system - the larger the loss of energy after a main shock, the better
is
the prediction of the next one. This is particularly clear for the
first
main shocks obtained when initiating the empty system, for which
predictabilty is very high. The astonishing accuracy observed in
\cite{logperiodic} for the Loma Prieta example might thus be due to
the
existence, in that case, of a very ``cool'' seismic system.

In addition, we  have demonstrated for the
first time the possible coexistence of self-organized criticality and
criticality. Up to now, they were considered as dual (mutually
exclusive)
modes of behavior: critical depinning occurs when the applied force
reaches a critical value beyond which the system moves globally,
while
self-organized criticality needs a slow driving velocity and
describes
the jerky steady-state of the system. The critical nature of our
large
cascades emerges from the interplay between the long-range
stress-stress
correlations of the self-organized critical state and the
hierarchical
geometrical structure: a given level of the hierarchical rupture is
like
a critical point to all the lower levels, albeit with a finite size.
The finite size effects are thus intrinsic to the process.


\begin{references}
\bibitem{foot} When computing this cumulative number, every breakdown
in  a cascade is counted as one event - this amounts  to putting
a  delay  of one time unit.
\end{references}
\end{document}